\documentclass{ws-p8-50x6-00}

\newcommand{\ii}{{\rm i}}
\newcommand{\de}{{\rm\,d}}
\newcommand{\e}{{\rm e}}

\newcommand{\im}{\,{\rm Im}\,}
\newcommand{\tr}{\mbox{Tr}\,}
\newcommand{\xbj}{x_{\scriptscriptstyle B}}                 
\renewcommand{\st}{{\scriptscriptstyle T}}

\newcommand{\g}{\gamma}
\newcommand{\sig}{\sigma}
\newcommand{\eps}{\epsilon}
\newcommand{\nn}{\nonumber}
\newcommand{\ovl}{\overline}

\newcommand{\pslash}{\rlap{/} p}
\newcommand{\kslash}{\rlap{/} k}

\begin{document}

\title{MEASURING TRANSVERSITY WITH T-ODD SINGLE PARTICLE PRODUCTION}

\author{A. Bacchetta}

\address{Department of Theoretical Physics, Faculty of Science, 
Free University \\
De Boelelaan 1081, NL-1081 HV Amsterdam, The Netherlands}

%%%%%%%%%%%%%%%%%%%%%%%%%%%%%%%%%%%%%%%%%%%%%%%%%%%%%%%%%%%%%%%%%%%%%%%

\maketitle

\abstracts{
The transversity distribution can be measured at leading-twist
in {\it single-particle inclusive} DIS, where it appears in
connection with chiral-odd fragmentation functions. 
Among these functions, we focus on the {\it time-reversal odd} ones.
They occur in fragmentation to unpolarized hadrons or polarized spin-one 
hadrons. 
Using a simple model, we show how these functions can be non-vanishing and we
discuss the asymmetries where they can be measured together with the
transversity distribution.}

%%%%%%%%%%%%%%%%%%%%%%%%%%%%%%%%%%%%%%%%%%%%%%%%%%%%%%%%%%%%%%%%%%%%%%%

The quark transversity distribution\cite{h1} $h_1$ 
 is an essential component in the description of 
the nucleon's spin.
Being chiral-odd, $h_1$ must be coupled to another chiral-odd function 
to appear in a physical observable.
In 1-particle-inclusive deep inelastic scattering, 
it can occur in transverse spin asymmetries at leading order in $1/Q$,
in connection with a fragmentation function.
When the final state hadron has no spin or spin 1, the involved chiral-odd 
fragmentation
functions belong to the class of the so-called time-reversal odd
(T-odd) functions. We are going to focus on these.

It is well established that T-odd fragmentation functions do not 
vanish in the case of two-hadron production\cite{marco}, 
but it is possible to show that this happens in one-hadron
production, too. 
In fact, the key feature to produce a T-odd function is the presence
of an interference between two scattering amplitudes
with different imaginary parts. 
In the case of two-hadron production, such an
interference can occurr, for instance, between the resonant and the 
non-resonant channels. In the case of one-hadron production, the simplest
expedient is to include one-loop corrections and to produce an interference
between the tree-level and the one-loop amplitudes. 

To show that this procedure indeed yelds non-zero T-odd functions, in a recent
article\cite{Bacchetta:2001di} we used a
simple model to calculate the Collins function\cite{Collins:1993kk}, which is
the archetype of a T-odd function.
Considering the fragmentation process $q^{\ast}(k) \to \pi(p) X$,
 we define the Collins function, 
which depends on the longitudinal momentum fraction $z_{\pi}$ of the pion and
the transverse momentum $k_{\st}$ of the quark, as 
\begin{equation} 
\frac{\eps_{\st}^{ij} k_{\st\,j}}{M_{\pi}}
				\, H_1^{\perp}(z_{\pi},k^2_{\st}) =
	 \left. \frac{1}{4z_{\pi}} \int \de k^+ \;
              \tr[ \Delta (k,p) \ii \sig^{i-}\g_5]
	\right|_{k^-=\frac{p^-}{z_{\pi}}},  
\label{e:col1}
\end{equation}
with $M_{\pi}$ denoting the pion mass, $\eps_{\st}^{ij} \equiv \eps^{ij-+}$
and $\Delta(k,p)$ denoting the quark-quark light-front correlation
function
% in Eq.~(\ref{e:col1}), omitting gauge
%links, takes the form
\begin{equation} 
\Delta(k,p)=\sum_X\, \int
        \frac{\de^{4}\xi}{(2\pi)^{4}}\,\e^{+\ii k \cdot \xi}\,
       \langle 0|
%\,{\cal L}[\xi,0;\mbox{path}]
\,\psi(\xi)|\pi, X\rangle \langle \pi, X|
             \ovl{\psi}(0)|0\rangle.    
\label{e:delta}
\end{equation} 

To describe the matrix elements in the correlation function, we use a 
pseudoscalar coupling between quarks and pions 
given by the interaction Lagrangian 
$
{\cal L}_I(x) =  \ii g \,\ovl{q} (x) \g_5 q(x) \pi(x) \,$.
This is clearly an oversimplified approach,
but the model contains the essential elements required for our discussion. 
In particular, it is time-reversal invariant.

\begin{figure}
	\centering
	\begin{tabular}{ccc}
	\epsfxsize=5.7cm
	\epsfbox{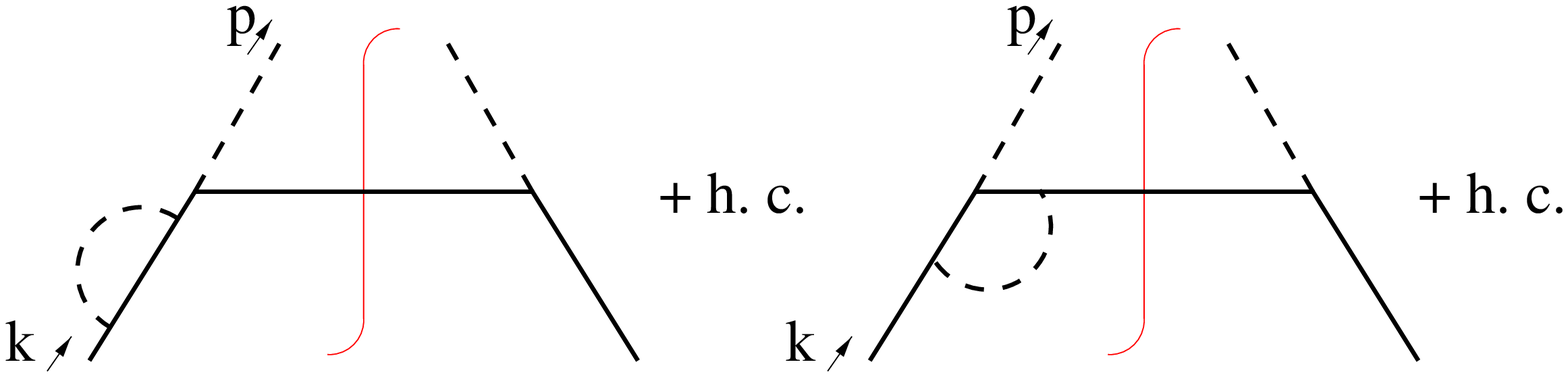} &&
	\epsfxsize=4.8cm
	\epsfbox{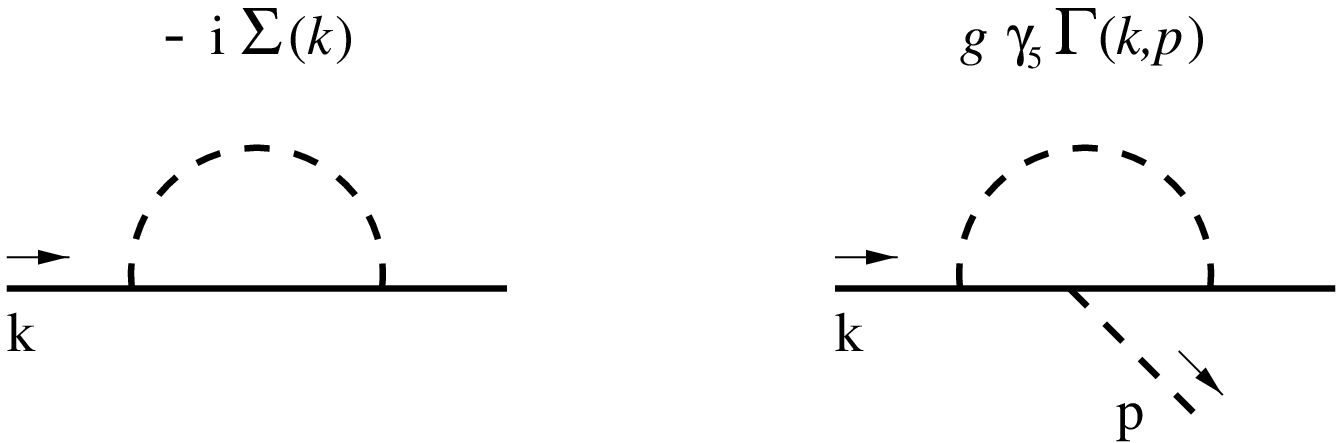} \\ \\
	(a) && (b)
	\end{tabular}
        \caption{One-loop corrections to the fragmentation of a quark
		into a pion relevant for the calculation of the Collins
		function (a), and details of the self-energy and vertex 
		corrections (b). }
        \label{f:1loop}
        \end{figure}

The one-loop diagrams contributing to the Collins function are 
shown in Fig.~\ref{f:1loop}~(a). 
In the computation, the relevant components to be included are the
self-energy correction and the vertex correction.
 They are sketched in
Fig.~\ref{f:1loop}~(b).
The functions $\Sigma (k)$ and $\Gamma (k,p)$ appearing in the diagrams can be
parametrized as
\begin{equation} 
\Sigma (k) \;=\; A\,\kslash + B\, m \,, \label{e:sig} \qquad \qquad 
\Gamma (k,p) \;=\; C + D\,\pslash + E\,\kslash + F\,\pslash \, \kslash .
\label{e:ver}
\end{equation}   

After calculating and summing the contributions of the four
relevant diagrams, we insert the 
resulting
correlation function in Eq.~(\ref{e:col1}) to obtain the result
\begin{eqnarray}  
\lefteqn{H_1^{\perp}(z_{\pi},k^2_{\st}) \;=}   \\
&&	\frac{g^2\,M_{\pi}}{8 \pi^3}\frac{1}{1-z_{\pi}}
	\left( \frac{m\, \im(A+B)}{(k^2 -m^2)^2}
	\mbox{}+ \frac{\im (D+E+mF)}{(k^2 -m^2)}\right) 
	\biggr|_{k^2=\frac{z_{\pi}}{1-z_{\pi}}k_{\st}^2
	+\frac{m^2}{1-z_{\pi}} +\frac{M_{\pi}^2}{z_{\pi}}}\,, \nn
\label{e:h1}
\end{eqnarray} 
where $m$ is the mass of the quark.

Thereby, we have shown that the Collins function can be generated by 
including one-loop
corrections in the fragmentation process. Though we
 performed an explicit calculation
using a simple model, we believe that the same procedure can produce T-odd
fragmentation functions in a much wider class of situations, with different
models and different particles in the final states. The inclusion of one-loop
corrections is therefore a simple and general way to produce T-odd {\it single
particle} fragmentation functions and gives a convincing indication that these
functions are non-zero.
Given these considerations, {\it one-particle inclusive} DIS stands out as a
promising process to investigate the transversity distribution of the nucleon.

The first case where the transversity distribution can be observed with T-odd
functions is the
production of a pion (or any unpolarized hadron) from a transversely polarized
target.
To single out the contribution containing the transversity distribution, 
it is required to measure the azimuthal asymmetry
\begin{eqnarray} 
\lefteqn{
\left\langle  \frac{|P_{\pi \perp}|}{z_{\pi} M_{\pi}}\;
   \sin{(\phi_{\pi}^\ell+\phi_S^\ell)} \right\rangle_{T}
%(\xbj,y,z_{\pi})
} 
\nn \\
&\equiv& 
{\int \de \phi_S^\ell  \de \phi_{\pi}^\ell\, \de |P_{\pi \perp}|\;
\frac{|P_{\pi \perp}|^2}{z_{\pi} M_{\pi}}
	\; \sin{(\phi_{\pi}^\ell+\phi_S^\ell)}\;
\frac{\de \sig^{\uparrow}-\de \sig^{\downarrow}}{\de(\xbj, 
		z_{\pi}, y, \phi_S^\ell, \phi_{\pi}^\ell, 
|P_{\pi \perp}|)}} \nn \\
&=& \frac{4\pi \alpha^2 s}{Q^4}\;\xbj\;\;|S_{T}|\;(1-y)\;
	{\sum_{a} e_a^2\,h_1^a (\xbj)\, H_{1}^{\perp (1) a} (z_{\rho})},
\end{eqnarray} 
where $P_{\pi \perp}$ is the perpendicular component of the pion
momentum, $y$ is the energy fraction of the scattered lepton,
 $\phi_S^\ell$ and $\phi_{\pi}^\ell$ are respectively the azimuthal
angles  of the target's transverse spin ($S_T$) and of the outgoing pion, 
with respect to the scattering plane. The summation runs over quark flavors
with charge $e_a$.

In the case of the production of spin-one hadrons\cite{Bacchetta:2000jk}, 
the extra information
about the polarization of the final state is
contained in the dependence of the cross section on the decay angles of the
decay products.
The
azimuthal asymmetries can in general be defined as \footnote{Although in the 
following we specialize the discussion on the $\rho$ meson, everything can be
applied also to higher mass vector mesons.}
\begin{eqnarray}
\left\langle W \right\rangle_{T}
&=& 
{\int \de \phi_S^\ell  \de \phi_{\rho}^\ell\, \de |P_{\rho \perp}| \, 
	\de \phi_{\pi}^\ell \;
W\;
\frac{\de \sig^{\uparrow}-\de \sig^{\downarrow}}{\de (\xbj, 
		z_{\rho},  y,  \phi_S^\ell,  \phi_{\rho}^\ell, 
|P_{\rho \perp}|,  \phi_{\pi}^\ell,  \theta_{\pi}) }},
\end{eqnarray}
where $W=W(\phi_{\rho}^\ell,|P_{\rho \perp}|,\phi_{\pi}^\ell,\theta_{\pi})$,
$\phi_{\pi}^{\ell}$ is  the azimuthal angle of one of the decay pions with
respect to the scattering plane and $\theta_{\pi}$ is the polar angle of the
same pion as measured in the $\rho$ meson rest-frame with respect to the
direction of the meson in the target rest-frame. 

The following asymmetries containing the transversity distribution can then be
obtained at leading order in $1/Q$:
\begin{eqnarray}  
\lefteqn{\left\langle\sin{(\phi_{\pi}^\ell+\phi_S^\ell)}\right\rangle_{T}}
	 \nn \\ &=&  
\frac{2\pi\alpha^2 s}{Q^4}\;\xbj\;\;|S_{T}|\;(1-y)\;|\sin{2\theta_{\pi}}|\;
	{\sum_{a} e_a^2\,h_1^a (\xbj)\, H_{1LT}^a (z_{\rho})}, 
	\label{e:H1LT}\\
\lefteqn{\left\langle \frac{|P_{\rho\,\perp}|}{2\,z_{\rho} M_{\rho}}\;
   \sin{(2\phi_{\pi}^\ell+\phi_S^\ell-\phi_{\rho}^\ell)}\right\rangle_{T}
	 } \nn \\ &=&  
\frac{4 \pi  \alpha^2 s}{Q^4}\;\xbj\;\;|S_{T}|\;(1-y)\;
	\sin^2{\theta_{\pi}}\;
	{\sum_{a} e_a^2\;h_1^a \,(\xbj)\; H_{1TT}^{(1)a}\, (z_{\rho})},
 \\
\lefteqn{\left\langle \frac{|P_{\rho\,\perp}|}{2\,z_{\rho} M_{\rho}}\;
	\sin{(\phi_S^\ell+\phi_{\rho}^\ell)}\right\rangle_{T}
	 } \nn \\ &=&  
\frac{4 \pi\alpha^2 s}{Q^4}\;\xbj\;\;|S_{T}|\;(1-y)\;
	\frac{\left(1-3 \cos^2{\theta_{\pi}}\right)}{2}
	\;\sum_{a} e_a^2\;
	h_1^a \,(\xbj)\; H_{1LL}^{\perp(1)a}\, (z_{\rho}), \\
\lefteqn{\left\langle \frac{|P_{\rho\,\perp}|^2}{2\,z^2_{\rho} M^2_{\rho}}\;
     \sin{(\phi_S^\ell+2\phi_{\rho}^\ell-\phi_{\pi}^\ell)}\right\rangle_{T}
	 } \nn \\ &=&  
\frac{4 \pi \alpha^2 s}{Q^4}\;\xbj\;\;|S_{T}|\;(1-y)\;
	|\sin{2\theta_{\pi}}|\;\sum_{a} e_a^2\;
	h_1^a \,(\xbj)\; H_{1LT}^{\perp(2)a}\, (z_{\rho}).
\end{eqnarray}
Note that the asymmetry of Eq.~(\ref{e:H1LT}) could represent a very
convenient way to measure the transversity distribution because it does not
require the perpendicular momentum of the outgoing hadron to be detected. 
It represents a specific
case of the more general two-pion production\cite{marco} and should exhibit 
a well 
known Breit-Wigner dependence on the invariant mass of the pion couple.

This work has been done in collaboration with R.~Kundu, A.~Metz and P.~Mulders
at the Free University of Amsterdam.

%%%%%%%%%%%%%%%%%%%%%%%%%%%%%%%%%%%%%%%%%%%%%%%%%%%%%%%%%%%%%%%%%%%%%%

\end{document}